\documentclass{sigchi-ext}
\usepackage[T1]{fontenc}
\usepackage{textcomp}
\usepackage[scaled=.92]{helvet} 
\usepackage{graphicx} 
\usepackage{balance}  
\usepackage{booktabs} 
\usepackage{ccicons}  
\usepackage{ragged2e} 
\usepackage{tabularx}

\def\plaintitle{Exploring Personality-Driven Personalization in XAI: Enhancing User Trust in Gameplay} 
\def\emptyauthor{}
\def\plainkeywords{Explainable AI; Personality; User preference}

\title{Exploring Personality-Driven Personalization in XAI: Enhancing User Trust in Gameplay}

\numberofauthors{6}
\author{%
  \alignauthor{%
    \textbf{Zhaoxin Li}\\
    \affaddr{Georgia Institute of Technology} \\
    \affaddr{Atlanta, GA 30332, USA} \\
    \email{zli3088@gatech.edu} }
    \vfil \alignauthor{%
    \textbf{Sophie Yang}\\
    \affaddr{Georgia Institute of Technology}\\
    \affaddr{Atlanta, GA 30332, USA}\\
    \email{soyang@gatech.edu} }
    \vfil \alignauthor{%
    \textbf{Shijie Wang}\\  
    \affaddr{Georgia Institute of Technology}\\
    \affaddr{Atlanta, GA 30332, USA}\\
    \email{shijie@gatech.edu} }
    }

\definecolor{linkColor}{RGB}{6,125,233}
\hypersetup{%
  pdftitle={\plaintitle},
  pdfauthor={\emptyauthor},
  pdfkeywords={\plainkeywords},
  bookmarksnumbered,
  pdfstartview={FitH},
  colorlinks,
  citecolor=black,
  filecolor=black,
  linkcolor=black,
  urlcolor=linkColor,
  breaklinks=true,
}

\begin{document}

\CopyrightYear{2020}
\setcopyright{rightsretained}
\conferenceinfo{CHI'20,}{April  25--30, 2020, Honolulu, HI, USA}
\isbn{978-1-4503-6819-3/20/04}
\doi{https://doi.org/10.1145/3334480.XXXXXXX}
\copyrightinfo{\acmcopyright}

\maketitle

\RaggedRight{} 

\begin{abstract}

  Tailoring XAI methods to individual needs is crucial for intuitive Human-AI interactions. While context and task goals are vital, factors like user personality traits could also influence method selection. Our study investigates using personality traits to predict user preferences among decision trees, texts, and factor graphs. We trained a Machine Learning model on responses to the Big Five personality test to predict preferences. Deploying these predicted preferences in a navigation game ($n=6$), we found users more receptive to personalized XAI recommendations, enhancing trust in the system. This underscores the significance of customization in XAI interfaces, impacting user engagement and confidence.

\end{abstract}




\begin{figure*}
    \centering
    \includegraphics[width= 0.85\textwidth]{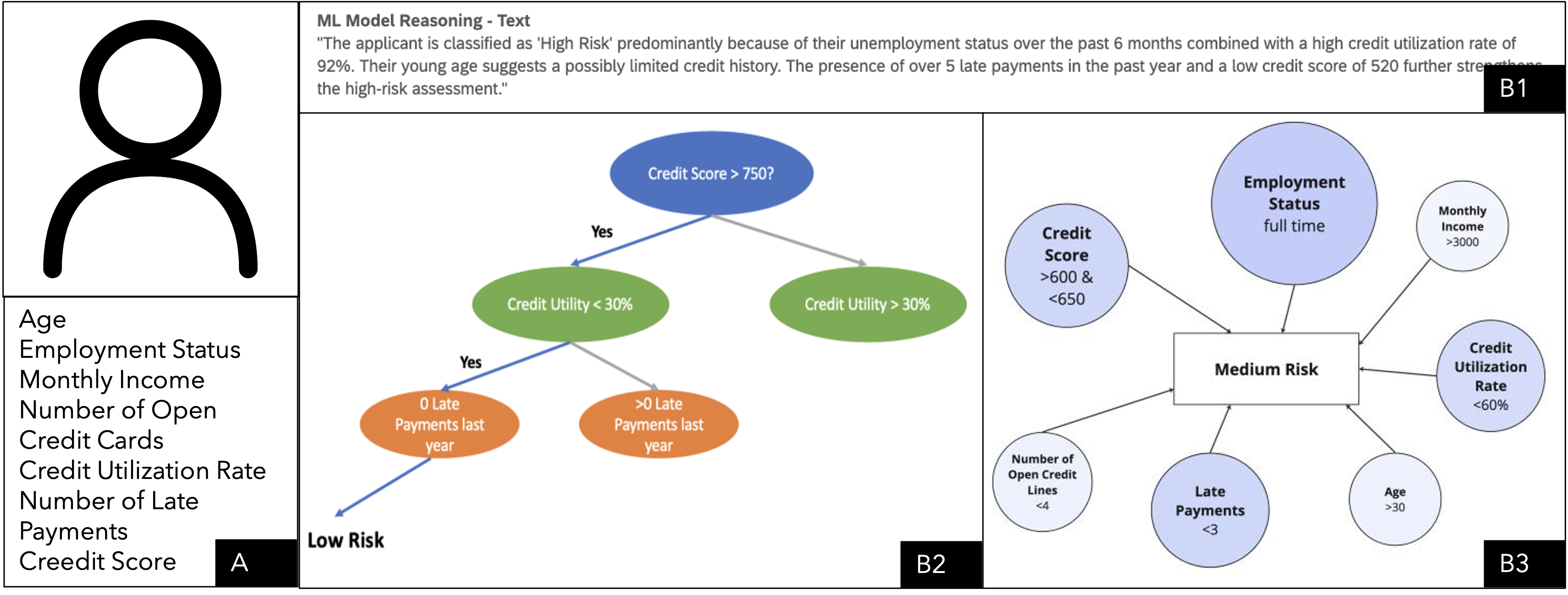}
    \caption{Capturing XAI preferences in Survey}
    \label{fig:gui}
\end{figure*}

\section{Introduction}
In recent years, AI has advanced significantly, with breakthroughs like GPT-4 in language modeling and improvements in computer vision and reinforcement learning. Alongside these advancements, there's been a growing emphasis on Explainable AI (XAI) to address concerns about transparency and ethics ~\cite{linardatos2020explainable}.


There is a shared consensus within the research community regarding the pivotal role of understanding users' mental models in determining the most suitable explanations for them. Bensi et al.~\cite{bensi2010individual} underscore the influence of personality traits on decision-making processes.  The findings highlight the significant impact of personalities on individuals' reasoning and decision-making. Building upon this notion, Hostetter et al.~\cite{hostetter2023xai} develop a pedagogical agent offering personalized explanations aligned with students' attitudes to enhance learning. Conati et al.~\cite{conati2021toward} also stress the significance of comprehending users' mental models to furnish appropriate explanations. In this study, we evaluated how understanding of personality traits might be leveraged to find the most intuitive XAI method to end users.


\begin{marginfigure}[-4cm]
  \begin{minipage}{\marginparwidth}
    \centering
    \includegraphics[width=0.9\marginparwidth]{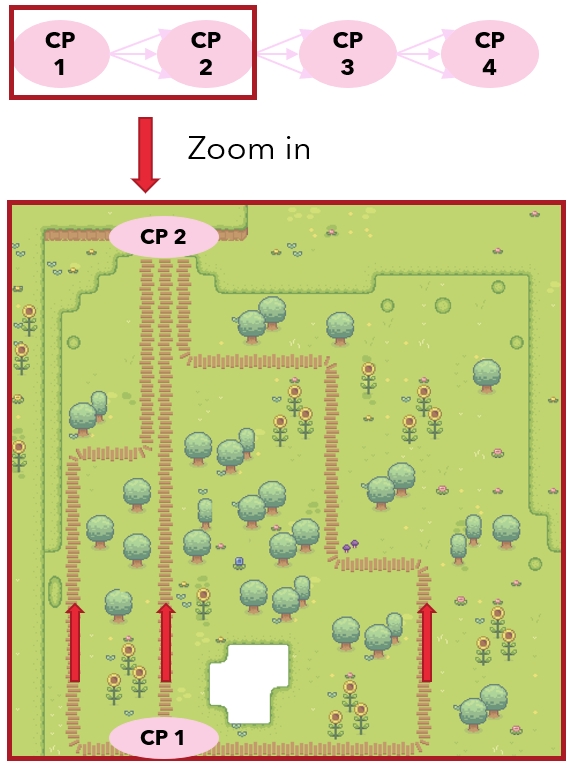}
    \caption{Overview of the game design.}~\label{fig:game_overview}
  \end{minipage}
\end{marginfigure}

\section{Experimental Design}

In this section, we describe our study design, research questions, metrics, domains, and experimental procedure.

\subsection{Survey Design}

To gather training data for our machine learning (ML) model, we devised a survey focusing on user personality traits and preferences for three XAI methods: decision trees~\cite{paleja2023interpretable}, textual explanations~\cite{celikyilmaz2020evaluation}, and factor graphs~\cite{loeliger2004introduction}. The survey comprises two sections: Part one evaluates personality traits using 20 questions based on the Big Five Factors of Personality. Part two involves an analytical task wherein users play the role of a loan manager at a bank, tasked with evaluating loan applicants' risk with the help of XAI recommendations. Upon completion, users are asked to rate their preferences for different XAI methods. We will be expanding on both parts in more detail below.

\subsubsection{The Big Five Personality}
The big five personality traits \cite{81cd5b1f64f14ce5979f22c14e6046fb}, developed in the 1980s in psychological trait theory, represent a grouping of five unique characteristics used to study personality. These characteristics include openness, neuroticism, agreeableness, extraversion, and conscientiousness. For this part of the survey, we employ the Mini-IPIP scales~\cite{donnellan2006mini} to measure these personality traits.

\subsubsection{XAI Preferences}
To assess participants' preferences among three XAI methods, we devised a scenario in which they acted as credit-scoring officers evaluating loan applications. Each participant reviewed background information on employment status, income, and credit utilization rate for each applicant. They then proceeded to part B, where they determined the credit risk level (high, medium, low) based on an XAI agent's recommendation and explanation. All 
participants are informed that the accuracy of the XAI agents are the same to avoid any bias in subsequent preference ratings.

\begin{marginfigure}
  \begin{minipage}{\marginparwidth}
    \centering
    \includegraphics[width=0.95\marginparwidth]{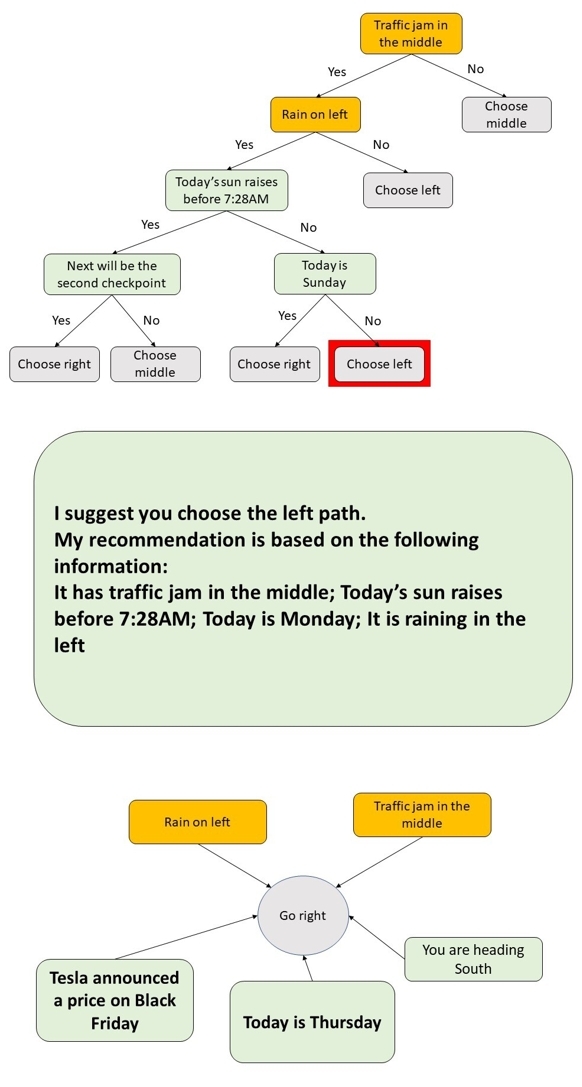}
    \caption{Three examples of XAI methods presented in the game (up: decision tree, middle: text, bottom: factor graph). The knowledge shared by both the agent and users is highlighted in orange in the decision tree and factor graph.}~\label{fig:XAI_methods}
  \end{minipage}
\end{marginfigure}

The survey presented three loan applications, each representing high, medium, or low risk profiles, paired with different explanation methods. Figure~\ref{fig:gui} shows an example of a loan application. High-risk profiles were paired with text explanations, medium-risk profiles with attribution graphics resembling factor graph, and low-risk profiles with decision trees. We opted not to randomize the pairing between risk levels and XAI methods because the explanations provided similar information, with the main difference being their organizational structure. Participants were asked to rank their preferences for the three XAI models.


\subsection{Machine Learning Model Design}
The machine learning model is designed to predict users' XAI preferences based on their personality data. It is trained on the data collected through our survey. We collected data from 11 respondents. A simple multi-layer perceptron is trained to map users' personality data to their XAI method preferences. The input dimension is 20, representing the personality traits, and the output dimension is 1, representing users' preferences for different XAI methods. At the end of the training both the training loss is converged to $< 1$. 





\subsection{The Game}
The aim of the game is to explore whether users are more likely to follow the XAI's recommendations when presented with their preferred method and less inclined when presented with their least favored method.
The high-level overview of the game is demonstrated in Figure \ref{fig:game_overview}. The game comprises four checkpoints, each connected by three paths. The layout of the game between each two checkpoints is adopted from~\cite{PyDewValley}. The user's objective is to select the path between each pair of checkpoints and guide the agent to the final destination (checkpoint 4) in the shortest amount of time possible. The longest route is the right path, requiring 15 minutes to complete. The middle path, being the shortest, takes 5 minutes to finish, while the left path has a duration of 10 minutes. 

The intuition of designing this game is that the expected time of complete each path between two checkpoint are the same. So participants are informs that it is always a event in the left path and middle path which adds uncertainty in the the completion time.  Specifically, the left path is consistently rainy, adding an extra time \(\delta_1 \sim \mathcal{N}(5, 5)\). On the other hand, traffic jams are inevitable on the middle path, introducing an additional time \(\delta_2 \sim \mathcal{N}(10, 5)\). Consequently, the total time for completing each path is determined by the following expressions:
 
\[t_{left} = (10 + \mathcal{N}(5, 5)) min\]
\[t_{middle} = (5 + \mathcal{N}(10, 5)) min\]
\[t_{right} = 15 min\]

As depicted above, the expected completion time for each path remains consistent. Consequently, the user's selection of the path is influenced by the recommendations provided by the XAI agent and their own judgment. At the first three checkpoints, an XAI agent suggests which path to take, employing methods such as text explanations, decision trees, or factor graphs (see Figure \ref{fig:XAI_methods}). XAI agent make its decision based on some additional information (e.g. temperature and time) that it thinks are relevant to making the decision. These methods are predetermined, and during gameplay, XAI decisions are randomly chosen for presentation. Users are informed that the XAI's accuracy is always \(1/3\) and need to decide whether to trust its suggestions. In other words, the XAI agent's recommendations resemble a random choice. This eliminates the possibility that participants may trust the agent based on its performance is higher than random chance, even if they don't particularly favor it. Ultimately, users' decisions to follow the XAI depend on their trust in the agent.


\subsection{User Study Design}

Six adults were invited to participate in the user study of playing the game. Three of the participants received their favorite XAI suggestion, while the other three received their least preferred XAI instruction. The study consisted of the following steps (a-d):

\begin{enumerate}
  \item[a)] Pre-questionnaire: Participants were first asked to answer a pre-survey questionnaire to answer the personality traits questions and indicate their age, gender, major, and learning style.
  \item[b)] Randomly assign the participant to either receiving his/her favourite XAI suggestion or least preferred XAI instruction.
  \item[c)] Let the participant play the game design in section 2.3. The chosen XAI method will be presented during the game process to provide participants with suggestions. The time used for finishing the game is recorded.
  \item[d)] A post-game survey is sent out to participants to collect data regarding participants’ perceived workload and usability of the XAI methods presented in the game.
\end{enumerate}

\subsection{Metrics}

During game play, we record the number of times users follow the XAI's suggestions, which can indicate the users' preference or trust in the presented XAI method. Furthermore, we also gather data on two metrics: perceived usability, which refers to users' subjective assessment of how easy and intuitive they find the system to use~\cite{lewis1992psychometric}; and trust in using the system, which pertains to users' confidence and reliance on the system's capabilities and integrity~\cite{jian2000foundations}.

\section{Results}

\subsection{Willingness of following XAI's suggestion}
During gameplay, we monitored the percentage of users' choices that aligned with the instructions provided by the XAI. The first group consisted of three participants who were assisted to their preferred XAI agent and the other group consisted of three participants who were introduced to their least preferred XAI methods. Each user controlled the agent through three checkpoints, resulting in a total of nine decisions made by users in each group. As illustrated in Figure~\ref{fig:follow_rate}, the group exposed to their preferred XAI method followed the XAI's suggestions 77.78\% of the time, while the group exposed to their least preferred XAI methods only followed the XAI agent's suggestions 44.44\% of the time. This demonstrates XAI's suggestions are more acceptable to users when the XAI method aligns with users mental model.




\subsection{Trust in using the system}
Following the gameplay, we collected participants' feedback on their trust in the system through a post-survey. Trust was assessed by measuring users' level of agreement with the statement "I trust the system." The results are displayed in Figure \ref{fig:bar_trust}. In the group exposed to their preferred XAI method, two users indicated "Agree," while one user responded with "Neutral." In the group assisted by their least favorite XAI methods, two users selected "Neutral," and one user chose "Disagree." This illustrates that presenting users with their preferred XAI methods helps foster trust in the system.

\subsection{Perceived usability}

\begin{marginfigure}[-3cm]
  \begin{minipage}{\marginparwidth}
    \centering
    \includegraphics[width=0.95\marginparwidth]{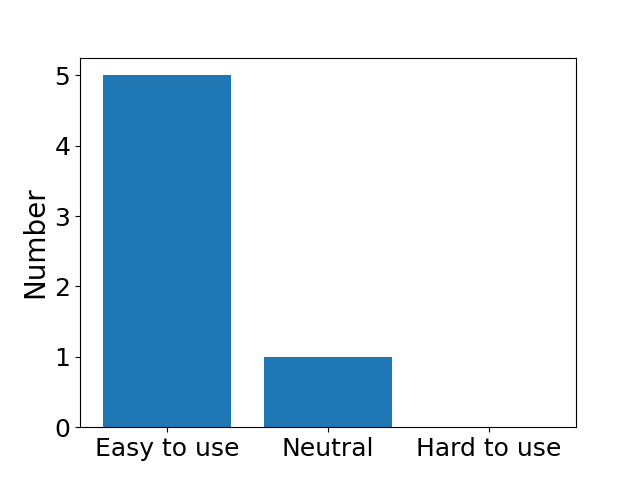}
    \caption{Users' perceived usability in using the system.}~\label{fig:bar_usability}
  \end{minipage}
\end{marginfigure}

Metric of perceived usability is measured by ask participants
the question of whether "I found the system very easy to use". As depicted in Figure~\ref{fig:bar_usability}, five out of six users responded affirmatively, indicating that they found it easy to use, while only one user remained neutral towards this statement. This suggests that our survey and game design necessitate minimal effort for users to comprehend and utilize.

\begin{marginfigure}[-25cm]
  \begin{minipage}{\marginparwidth}
    \centering
    \includegraphics[width=0.95\marginparwidth]{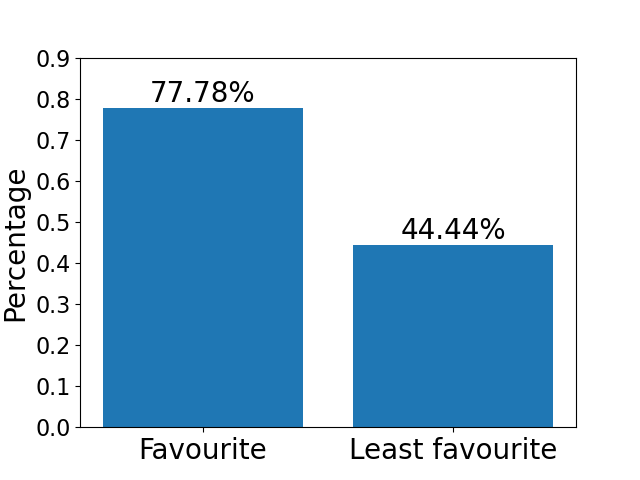}
    \caption{The percentage of users' decisions that align with XAI's instruction.}~\label{fig:follow_rate}
  \end{minipage}
\end{marginfigure}

\begin{marginfigure}[-13cm]
  \begin{minipage}{\marginparwidth}
    \centering
    \includegraphics[width=0.95\marginparwidth]{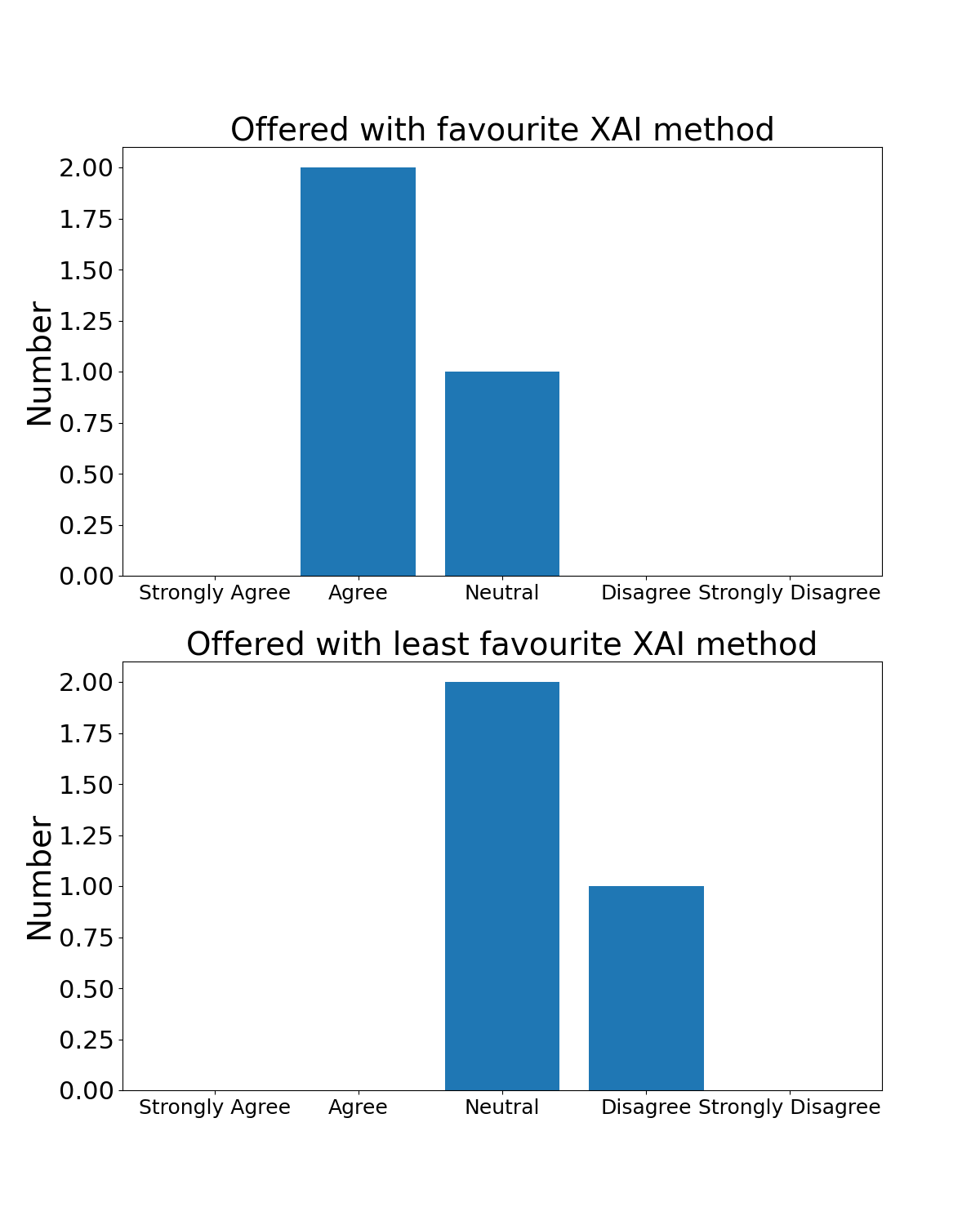}
    \caption{Users' trust in using the system.}~\label{fig:bar_trust}
  \end{minipage}
\end{marginfigure}

In summary, users are more likely to follow XAI's suggestions when suggestions align with their preferred XAI methods, leading to higher trust in the system (Figures \ref{fig:follow_rate} and \ref{fig:bar_trust}). Additionally, participants consistently found the system easy to use (Figure \ref{fig:bar_usability}). These findings collectively highlight the importance of aligning XAI recommendations with user preferences to foster trust and usability.

\section{Conclusion}
Our research highlights the importance of personalizing Explainable Artificial Intelligence (XAI) methods based on individual personality traits. By predicting users' preferences, we tailored XAI methods in a navigation game, leading to increased willingness to follow suggestions. This underscores the nuanced relationship between personality and XAI preferences. Participants' feedback reinforced the positive impact of personalized XAI, with higher trust levels observed for favorite methods. This implies that in the design of XAI systems, developers can take users' personality traits into account and ensure that the XAI method aligns with the user's preferences. This alignment can help cultivate trust and affinity towards the system. In essence, our study unveils the intricate relationship between user characteristics, XAI preferences, and system trust, paving the way for future developments in personalized XAI systems.




\balance{} 

\bibliographystyle{SIGCHI-Reference-Format}
\bibliography{sample}

\end{document}